\documentclass[twocolumn,pra,showpacs]{revtex4}
\usepackage{amssymb,amsmath,amsfonts,bm}
\usepackage{graphicx}
\usepackage{color}
\renewcommand{\k}{\mathbf{k}}

\begin{document}

\date{\today}

\title{Phase transitions in optical turbulence}

\author{Natalia Vladimirova$^1$,
Stanislav Derevyanko$^2$ and Gregory Falkovich$^3$}
\affiliation{$^1$University of New Mexico,
Department of Mathematics and Statistics\\
$^2$Nonlinearity and Complexity Research Group, Aston University, Birmingham UK\\
$^3$Weizmann Institute of Science, Rehovot 76100 Israel}

\begin{abstract}
  We consider turbulence in the Gross-Pitaevsky model and study
  the creation of a coherent condensate via an inverse cascade
  originated at small scales. The growth of the condensate leads to a
  spontaneous breakdown of  symmetries of small-scale over-condensate
  fluctuations: first, statistical isotropy is broken, then series of
  phase transitions mark the change of symmetry from the two-fold to three-fold
  to four-fold. At the highest condensate level reached, we observe a short-range
  positional and long-range orientational order (similar to a hexatic
  phase in the condensed matter physics). In other words, the longer one pumps the system the more
  ordered it becomes. We show that these phase transitions happen when the driving term corresponds to an instability (i.e. it is multiplicative in  the ${\bf k}$-space) but not when the system is pumped by a random force. Thus we demonstrate for the first time  non-universality of the
  inverse-cascade turbulence. We also describe anisotropic spectral
  flux flows in ${\bf k}$-space, anomalous correlations of
  fluctuations and collective oscillations of turbulence-condensate
  system.
\end{abstract}


\maketitle

\section{Introduction}
Probably the most unexpected discovery made in the study of turbulence is
an inverse cascade.  Flying at the face of an intuitive picture of
turbulence as the process of fragmentation and dissipation, inverse
cascade means self-organization, i.e. appearance of large-scale
motions out of a small-scale noise. The process of inverse cascade
culminates in the creation of spectral condensate i.e. a mode
coherent across the whole system. At this point turbulence shares many
properties with quantum physics as displaying both fluctuations and
coherence. This closeness shows perhaps most vividly in the
studies on non-equilibrium states within the framework of Nonlinear
Schr\"odinger (or Gross-Pitaevsky) Equation, called here NSE. In the
realm of classical physics, this model describes spectrally narrow
distribution of nonlinear waves, the respective far-from-equilibrium
states are called optical turbulence \cite{ZLF,Naz,Fal}. For quantum physics, this is the
simplest model of locally interacting bosons. The Hamiltonian and the
canonical NSE are as follows
\begin{eqnarray}
&&{\cal H}=
  \int\Bigl(|\nabla\psi|^2+|\psi|^4/2\Bigr)\,d^2r\,,\label{Ham}\\
&&i\psi_t=\nabla^2{\cal H}/\delta \psi^*=
  -\Delta\psi+|\psi|^2\psi\ .\label{NSE}
\end{eqnarray}

To provide for a turbulent inverse cascade, we add medium-scale pumping and
small-scale dissipation and solve the resulting equation numerically with a
record spatial and temporal resolution. We address the central issue in the
non-equilibrium physics, i.e. that of universality: to what extent the
properties of a far-from-equilibrium state depend on the excitation
mechanism? Theoreticians prefer to pump a turbulent system by a random Gaussian force,
which practically never can be met in real systems where waves usually
appear as a result of some instability. Here we consider both ways to
excite waves (additive random forcing and instability) and show that they produce
similar turbulence only when the condensate is weak.  We show that the
instability-excited system undergoes a series of phase transitions to
the states with different symmetries while the force-driven system
does not.

Understanding the interaction of turbulence and a coherent flow is
an important problem in turbulence studies in fluid
mechanics and beyond, both from fundamental and practical
perspectives. In fluids, coherent flows  are system-size vortices or
zonal flows of different profiles~\cite{Shats,CKL}, which are known to
diminish the turbulence level, change its nature
and make its statistics more non-Gaussian~\cite{Shats,NP}. Here we consider arguably
the simplest case of a turbulent system with a condensate: the coherent part
is expected to be a constant field and turbulence to consist of weakly
interacting waves. We carry our simulations well past time when all vortices, i.e. holes
in the condensate, disappear (due to annihilations~\cite{no}) and observe that the coherent part contains
a small but important spatial structure and the turbulence is not
effectively weak.  As a result, this simple system demonstrates
unexpectedly rich behavior with novel features never before observed
in turbulence systems.

\section{Theoretical considerations}

Let us denote $N=\overline{|\psi|^2}$ and $N_0=|\overline{\psi}|^2$.
In what follows the overline means average over space, and the averages over time we shall denote
by angular brackets.  The total number of waves with nonzero wave numbers is
$N-N_0=n=\int |\psi_k|^2d^2k$.

The simplest condensate is a spatially-uniform field, $A=\sqrt{N_0}\exp(-iN_0t)$,
which is an exact solution of (\ref{NSE}). Small over-condensate
fluctuations satisfy
\begin{equation}i\dot\psi_k=(k^2+2N_0)\psi_k+A^2\psi_{-k}^*.
\label{linear}
\end{equation}
This equation gives  the Bogolyubov dispersion relation,
\begin{equation}
  \Omega_k^2=2N_0k^2+k^4 \ ,
  \label{Bog}
\end{equation} for a pair of counter-propagating waves
$\psi_k\propto \exp(-iN_0t-i\Omega_kt)$,  $\psi_{-k}^*\propto \exp(iN_0t-i\Omega_kt)$.


\subsection{Turbulence with the condensate}

As the condensate grows, the dispersion relation (\ref{Bog})
approaches the acoustic one with the sound velocity
$c=\sqrt{2N_0}$. As known, acoustic waves running at the same
direction interact strongly (producing shocks). On the other hand, the
effective matrix element of the three-phonon interaction behaves as
$|V_{123}|^2\simeq k_1k_2k_3/ N_0^{1/2}$ (see e.g. \cite{Opt}). One can
estimate the mean value of the cubic term in the Hamiltonian using the
weak-turbulence approximation
\cite{ZLF}:
\begin{eqnarray}
\langle{\cal H}_3\rangle\!&=&\sum_{{\bf k}_1,{\bf k}_2,{\bf k}_3}V_{123}
\langle \psi_{k_1}\psi_{k_2}\psi_{k_3}^*\rangle
\delta(\k_1+\k_2-\k_3)\nonumber\\\!&\simeq&\!\!\!\!\!\!\sum_{{\bf k}_1,{\bf k}_2,{\bf k}_3}\!\!\!\!\! |V_{123}|^2n_1n_2\, \delta({\bf k}_1+{\bf k}_2-{\bf k}_3)\delta(\Omega_1+\Omega_2-\Omega_3)\nonumber\\
\!&\simeq& {|V|^2n^2c\over k^3}{k\over c}\simeq {n^2k\over N_0^{1/2}}
\label{estimate}\ .
\end{eqnarray}
 Here we have introduced the mean
spectral density of waves also called the normal correlation function
$n_k=\langle |\psi_k|^2\rangle$. The factor ${k/ c}$ in (\ref{estimate}) is
the effective angle of interaction that can be estimated considering
$\Omega_k\approx c|k|+k^3/2c\approx ck+ck_\perp^2/2k+k^3/2c$ and
comparing diffraction and dispersion corrections: $k_\perp/k\simeq
k/c$.  We then estimate $\langle {\cal H}_2\rangle \simeq
ck n$ and obtain the effective nonlinearity parameter for
over-condensate phonons as the ratio:
\begin{eqnarray}
{{\cal H}_3\over{\cal H}_2}\simeq {n\over N_0}\ .\label{estim1}
\end{eqnarray}
This suggests that when the number of waves with zero momentum outgrows
the number of waves with nonzero momentum, nonlinear interaction must be
getting effectively weak despite the fact that the nonlinear term in
the Hamiltonian is dominant. The estimate (\ref{estim1}) explains the
observation made in \cite{DF} that the statistics of the fluctuations
is non-Gaussian at low condensate values but is getting closer to
Gaussian, as the condensate grows.

Formula (\ref{estimate}) allows also to estimate the typical rate of
a local nonlinear interaction (i.e. for waves with comparable $k$) as
$1/t_{nl}(k)\simeq |V|^2 n_k k^2(c/k^3)(k/c)\simeq n_kk/ N_0^{1/2}$.  The rate of nonlinear interaction
$1/t_{nl}$ decreases as the condensate grows. Therefore, nonlocal
interactions must play more and more important role. Here we define
the coherent part as the field filtered at the frequency $N$ and show
that it is not a constant but has a rather elaborate spatial
structure. Interaction of turbulence and the coherent part (nonlocal
in ${\bf k}$-space) is responsible for the set of new phenomena described
below.

One may think that an effective weak nonlinearity expressed by
(\ref{estim1}) allows one to describe the over-condensate turbulence
by the weak turbulence theory i.e. by a closed kinetic equation
written for the one-time (normal) correlation function $n_k$. We will see
here that this is not necessarily the case.  The crucial point is that
validity of weak turbulence approach requires not only weak
nonlinearity but random phases as well. We show below (analytically)
that condensate imposes phase coherence, the most straightforward
manifestation of which is the existence of anomalous correlation
functions and the phase coherence of the $k,-k$-pairs with the
condensate.  We then describe the series of phase
transitions (discovered by the direct numerical simulations) that show
evolution towards more and more ordered state as the condensate grows;
such evolution cannot be described by a kinetic equation, which
satisfies H-theorem and can describe only entropy growth.

\subsection{Anomalous correlations}
\label{sec:anom-cor}

Let us now show that an important element of turbulence against the
background of a condensate must be anomalous correlations, i.e. phase
coherence between waves running in opposite directions (seems to have
been overlooked so far).  Indeed, the existence of the condensate must
produce the anomalous correlation function $\langle
\psi_k\psi_{-k}\rangle \propto \exp(-2iN_0t)$. To look into the
effective dynamics of both the number of waves and the condensate, let
us assume for simplicity that only two contra-propagating waves with
amplitudes $\psi_{\pm k}=\sqrt{n_{\pm k}}\exp(-iN_0t+i\,\phi_{\pm k})$
interact with the condensate and expand the Hamiltonian (\ref{Ham}) up
to the terms quadratic in $\psi_{\pm k}$.  That gives the equations:
\begin{equation}
\begin{split}
{dn_k\over dt}& =-\frac{1}{2}\,{dN_0\over dt}=-B=2N_0\sqrt{n_kn_{-k}}\sin(2\phi_0-\phi_k-\phi_{-k})\,,\\
{dB\over dt} & =4(2n_k+2n_{-k}-N_0-k^2)C+8N_0n_kn_{-k}\nonumber\\&-2N_0^2(n_k+n_{-k}) \,,
\\
{dC\over dt} & = (N_0+k^2-2n_k-2n_{-k})B
\end{split}
\label{phase2}
\end{equation}
where we have denoted
$ A^2\psi_k^*\psi_{-k}^* =C-iB/2$.  We see that the steady state
corresponds to $B=0$ and $C=-N_0\sqrt{n_kn_{-k}}$ that is to $2\phi_0-\phi_k-\phi_{-k}=\pi$ (and  $n_k=n_{-k}=k^2/2$).  Alternatively, one
calculates the time derivative of the Hamiltonian, quadratic in
$A,\psi_{\pm k}$: $d{\cal H}_2/dt\approx
4N_0Im(A^24\psi_{-k}^*\psi_k^*) =
N_0^2\sqrt{n_kn_{-k}}\sin(2\phi_0-\phi_k-\phi_{-k})$, which is the
energy flux from the condensate to the wave pair with $k,-k$.  The
flux turns to zero and the steady state is established when the phase of the pair is exactly opposite to the phase of $A^2$, which plays the role of an
effective pumping.


\subsection{Anisotropy of turbulence}\label{sec:aniz}

At the beginning, waves in the pumping shell are pumped isotropically. Eventually, the inverse
cascade reaches lowest modes, which have  the four-fold symmetry of the box. Let us now discuss
what possible anisotropy may turbulence spectra acquire after the
condensate appears.  Just looking at the Bogolyubov dispersion
relation (\ref{Bog}) one may suggest the following scenario: spatial
modulation of the condensate intensity $N_0$ means respective
modulation of the sound velocity for shorter waves. Standing waves
corresponding to the lowest modes of the box provide such
modulation. Regions of low sound velocity would act as waveguides with
short waves moving predominantly along the minima of $N_0$. That
suggests that for sufficiently high $N_0$, when (\ref{Bog}) is close
to linear, the angular maxima
at high $k$ would correspond to minima  at
low $k$ for a hypothetical anisotropic  spectrum. In other words, acoustic waves interact effectively when they
are collinear, so the maximum of long waves at some angle effectively
removes short waves from the pumping region at that angle. That
consideration suggests a simple picture of possible anisotropy: lowest
modes have the symmetry of the square box and impose that
symmetry on the small-scale turbulence (turned by $\pi/4$). The main
result of this work is that this is not the case.

\section{Numerical procedure}
\label{sec:numerics}

Our numerical simulations are set up similarly to~\cite{DF}.
We use 4th order fully dealiased split-step method~\cite{agrawal,yoshida}
to solve Eq.~(\ref{NSE}) in the presence of forcing,
\begin{equation}
i\,\psi_t =  -\nabla^2 \, \psi + |\psi|^2 \,\psi + i \hat{\gamma} \,\psi.
\label{NLSE}
\end{equation}
The forcing consists of  large scale pumping and small scale damping.
The pumping-attenuation operator
$\hat{\gamma}$ is local in the ${\bf k}$-space and is given by
$\gamma(k)=\gamma_p(k)-\gamma_d(k)$ where the pumping is non-zero only
within the shell~\cite{DF},
\begin{equation*}
  \gamma_p(k)=\alpha\,\sqrt{(k^2-k_l^2)(k_r^2-k^2)}
  \quad \mathrm{at} \quad k_l<k<k_r
  \label{pumping}
\end{equation*}
and the damping is of Landau-type,
\begin{equation*}
  \gamma_d(k)=\beta\,k^2\,h(k/k_d),
  \label{damping}
\end{equation*}
where function $h(x)$ is defined as
\begin{equation*}
   h(x)= \left\{ \begin{array}{cc}
       (6\,x^5)^{-1}\,\exp\left[5\,(1-x^{-2})\right], & x \leq 1; \\
       & \\
       1-(5/6)\,\exp \left[(1/2)\,(1-x^2)\right], & x>1.
   \end{array}\right.
   \label{eq-h*}
\end{equation*}
The pumping shell $[k_l,k_r]$ is relatively narrow and
occupies the middle of the computational domain. Note that we do not
suppress long wave excitations (no damping is present at $k=0$) so, as
expected, the condensate starts to form early on and the total number
of waves begins to grow in time.  As in~\cite{DF}, the initial condition is
given by thermal equilibrium spectrum, $n_k=T/(k^2 +\mu)$, with random
phases.

\begin{figure*}
\includegraphics[width =0.48\textwidth]{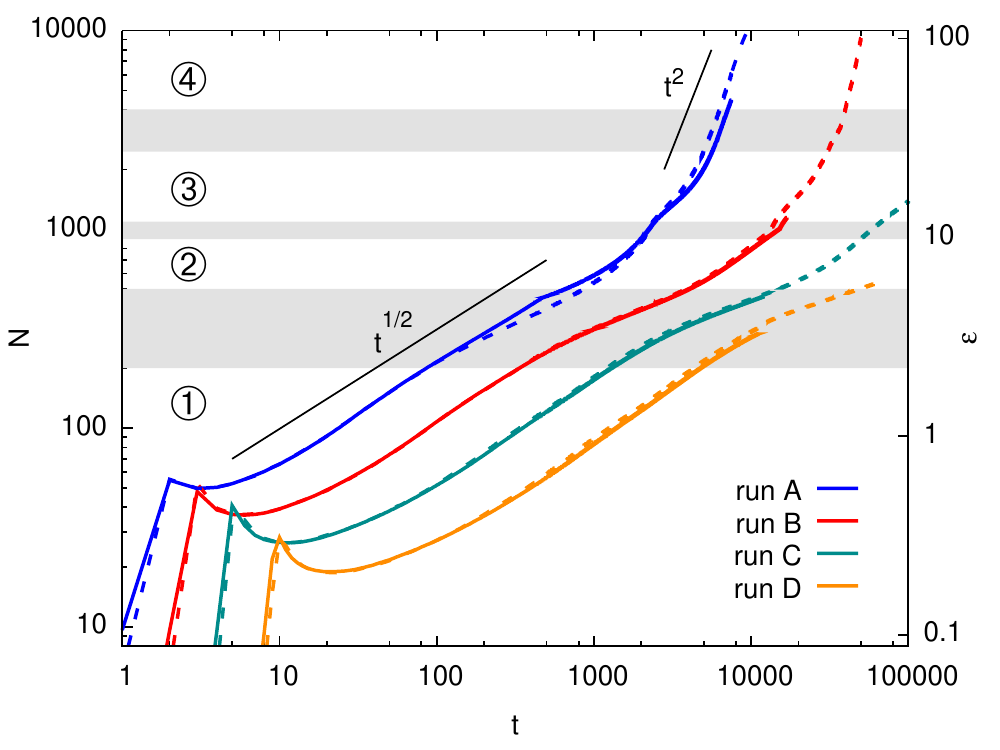}
\hfill
\includegraphics[width =0.48\textwidth]{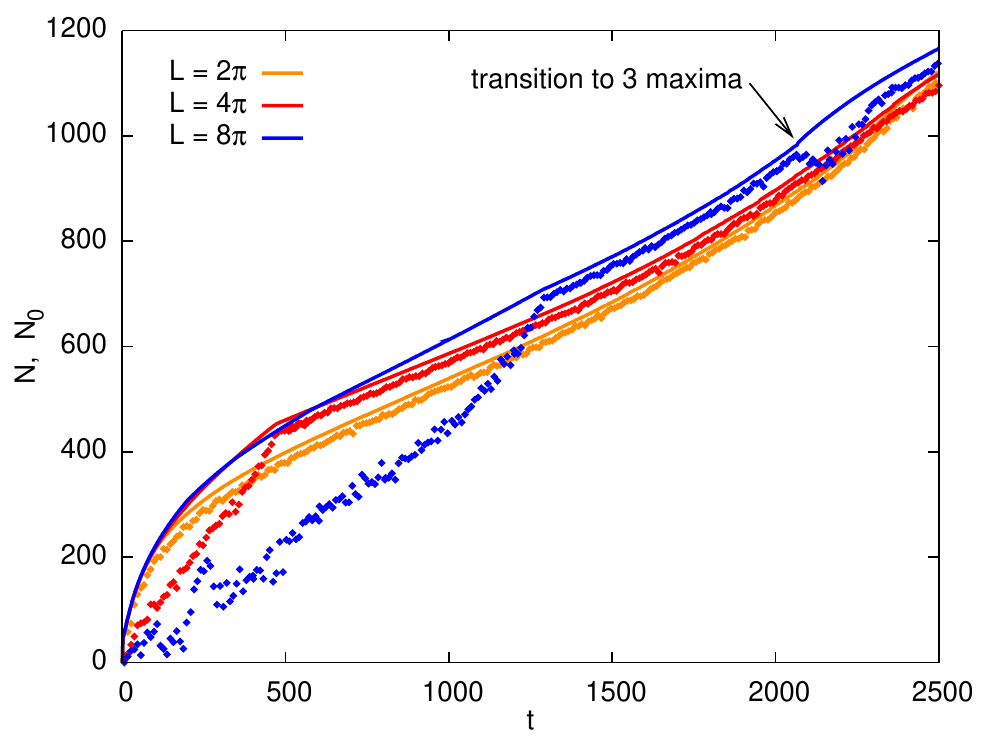}
\caption{Left: Number of waves as a function of time in
$L=2\pi$ (dashed lines) and  $L=4\pi$ (solid lines) simulations,
also rescaled as $\epsilon$.
 Gray regions correspond to the transition between regimes;
circled labels refer to the number of jets (petals).
Right:
Appearance of condensate in simulations with different box sizes in Run A.
The condensate value $N_0$ is shown by dots,
next to the lines representing $N$.
The initial linear growth of the condensate at strong forcing and
large boxes is accompanied by creation and destruction of vortices. When vortices die out,
$N_0$ catches up with $N$.
}
\label{fig:numwaves_log}
\label{fig:nk_scaled}
\label{fig:numwaves}
\end{figure*}

The values for simulation parameters are shown in
Table~\ref{tab:parameters}.  Note that the size of the box, $L$, increases
with resolution, while forcing remains the same. As a result the
number of waves per unit area, $N$, evolves in a similar way in simulations at
different box sizes, see Figure~\ref{fig:numwaves_log}a. Unless specified,
the results are shown below for the run A.

\begin{table}
\begin{tabular}{| l | c | c | c |}
\hline
box size $L$  & $2\pi$ & $4\pi$ &  $8\pi$ \\
\hline
resolution & $128$  & $256$ & $512$ \\
timestep   & $10^{-4}$ & $10^{-4}$ & $10^{-4}$ \\
$\sqrt{\mu}$ & 0.15 & 0.3 & 0.6 \\
initial $NL^2$ & 15 & 60 & 240 \\
forcing: $k_l$   & 28 & 56 & 112 \\
forcing: $k_r$   & 32 & 64 & 128 \\
forcing: $k_d$   & 42 & 84 & 168 \\
forcing: $\beta$ & $10 \alpha$  & $10 \alpha$  & $10 \alpha$ \\
\hline
forcing: $\alpha$ &&&\\
run A & $6.40\times 10^{-2}$ & $1.60\times 10^{-2}$ & $4.00\times 10^{-3}$ \\
run B & $3.20\times 10^{-2}$ & $8.00\times 10^{-3}$ & $2.00\times 10^{-3}$ \\
run C & $1.60\times 10^{-2}$ & $4.00\times 10^{-3}$ & $1.00\times 10^{-3}$ \\
run D & $8.00\times 10^{-3}$ & $2.00\times 10^{-3}$ & $5.00\times 10^{-4}$ \\
\hline
\end{tabular}
\caption{ Note: (a) resolution is shown excluding dealiased modes;
(b) the timestep is reduced to $5\times10^{-5}$ when $\max|\psi|^2$
reaches $2\times10^3$,
and to $2\times10^{-5}$ when $\max|\psi|^2$ reaches $4\times10^3$.}
\label{tab:parameters}
\end{table}

To collect better statistics, we occasionally stop the simulation and
restart it with uniform friction which stabilizes the condensate
growth and leads to a steady state. In this case, the forcing is
replaced by $\tilde{\gamma} = \gamma - b$, where~$b$ is constant
determined empirically to stabilize~$N$ at a given
level.

To study the structure of the coherent mode and
fluctuations separately, we applied the time-spectral filtering used by
Nazarenko and Onorato~\cite{no}.  After the condensate appears, the
frequency spectrum has a sharp peak at $\omega_0=N$.  We select a time
window containing many $2\pi/\omega_0$ periods (typically 100) and store
the temporal evolution of the $\psi(r,t)$ with 12-24 time-slices per
period.  This data is transformed to $\omega$-$k$ space, where 
the coherent mode ---
slice corresponding to $\omega_0$ --- is separated.
The rest of the frequency spectrum
we treat as fluctuations.

We use as a measure of nonlinearity, the dimensionless parameter
$\epsilon=N\lambda^2/4$ where $\lambda$ is the pumping wavelength,
which is not only the smallest scale in the inverse cascade but also
the smallest scale in the system since dissipation dominates for the
scales less than $\lambda$.  All our simulations are done with the
same $\lambda \approx 2\pi/30$, so $\epsilon$ also has the meaning of
rescaled number of waves, $\epsilon \approx 0.01 N$.

\section{Results and discussion}\label{sec:result}

\subsection{Condensate growth}

The nonlinear interaction conserves the total number of waves so the latter grows only due
to net pumping: $dN(t)/dt=\Gamma(t)=\int \gamma_kn_k(t)\,d^2k$.  A simple
expectation would be that $N(t)$ must initially
grow exponentially as a result of linear instability, then
nonlinearity saturates this growth.  After $n_k$ stabilizes in the
pumping region, one expects a linear law $N=\Gamma t$. The first
study of the wave number growth was done in~\cite{DF} where it was
observed that nonlinearity indeed slows it down (and for a while even reverses) after the initial exponential increase. The following assertions
were then made in ~\cite{DF}: (a) the first nonlinear stage of growth follows
the law $N\propto\sqrt{t}$, which means that the occupation numbers
in the pumping shell decrease as $n_k\propto 1/\sqrt{t}$; (b) the
square-root regime starts when the number of waves at the condensate
is of order of the total number of waves; (c) the square-root
regime ends when $\epsilon\simeq1$ i.e. the correlation scale
$N_0^{-1/2}$ is approaching the pumping scale; (d) after that, $n_k$ stabilizes and $N=\Gamma t$ with time-independent $\Gamma$. As seen
from the Figure~\ref{fig:numwaves_log} (left panel) (produced at the resolutions much
higher and on time intervals much longer than those of~\cite{DF}), the
evolution, which we observe, does not follow this simple scenario.  At earlier times, the
growth on average can be crudely approximated as $\sqrt{t}$.  In some
simulations, we observed a transitional linear regime, as
in~\cite{DF}. It can be seen on a linear scale, as, for instance, the
section $500<t<1800$ for $L=8\pi$ curve in
Figure~\ref{fig:numwaves_log} (right panel). At later times, the growth rate of the
condensate accelerates even faster than linear, with possible
transition to the ultimate regime $N\propto t^2$ regime hypothesized
by Zakharov and Nazarenko~\cite{ZN}. The growth $N(t)$ is determined by the number of waves in the pumping shell, and we shall see in the next section that this number evolves in a complicated way due to phase transitions. We found the law of this growth
sensitive to different parameters and were not able to distinguish any
definite scaling.  That subject definitely requires extensive future
studies.


\subsection{Phase transitions}

\begin{figure*}
\includegraphics[width=0.98\textwidth]{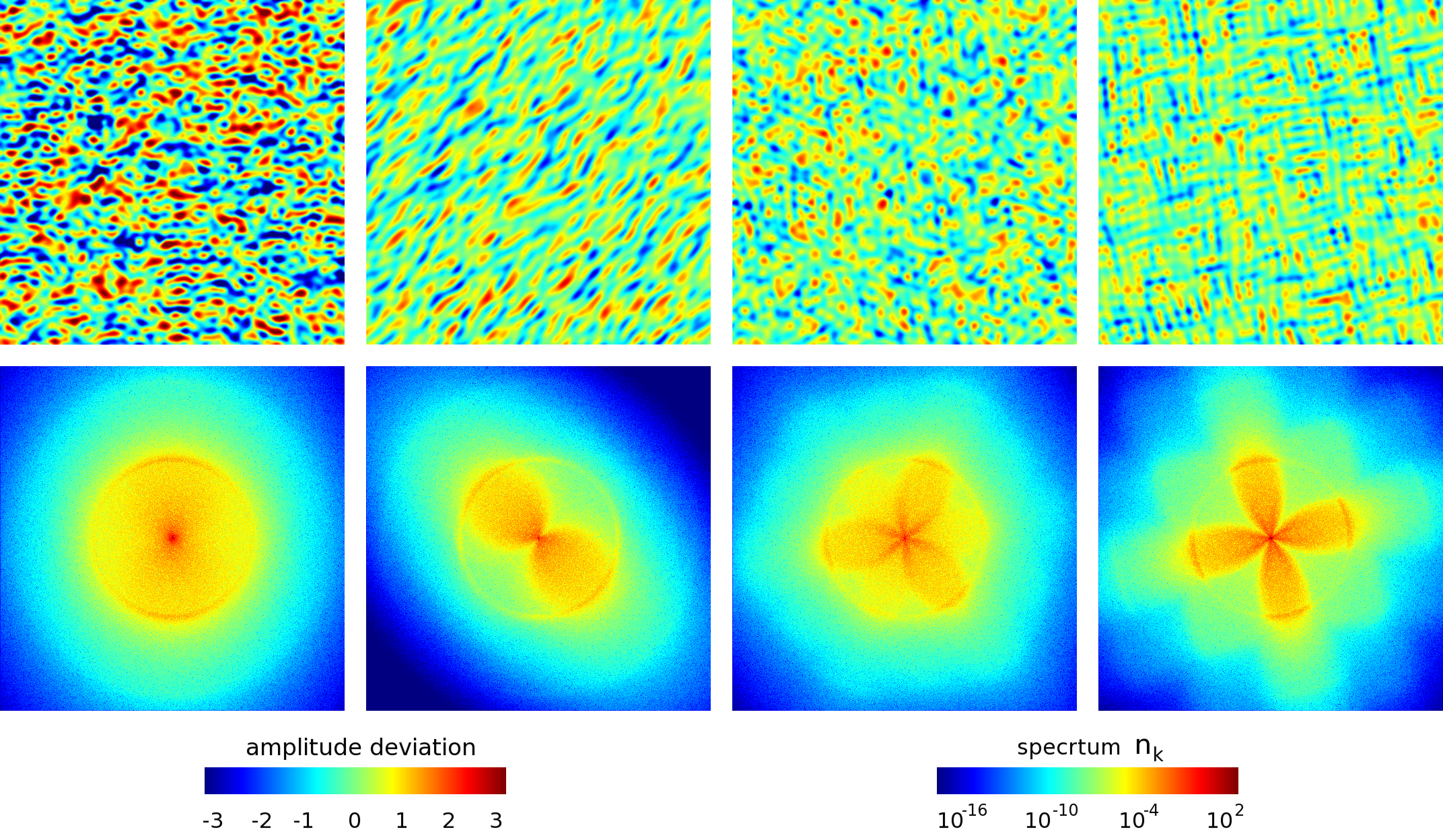}
\caption{
Amplitude deviation from average in a 
$2\pi \times 2\pi$ part of the computational
domain (above) and $n_k$ spectrum (below) in $L=8\pi$ simulation.
From left to right images correspond
to $t=100$, $1500$, $2500$, and $7500$ ($N=219$, 771, 1166, 4202,
$\epsilon = 2.40$, 8.45, 12.77, 46.08).
}
\label{fig:zoo}
\end{figure*}

\begin{table*}
\begin{tabular}{| c | c || c | c | c || c | c | c |}
\hline
   \multicolumn{2}{|c||}{transition} &
   \multicolumn{3}{|c||}{two-fold to three-fold } &
   \multicolumn{3}{|c|}{three-fold to four-fold} \\
\hline
 run & $L$ &
   \makebox[22mm]{$t$} & \makebox[20mm]{$N$} & \makebox[22mm]{$\epsilon$} &
   \makebox[22mm]{$t$} & \makebox[20mm]{$N$} & \makebox[22mm]{$\epsilon$} \\
\hline
A     & $2\pi$ & $2150 \pm 50$ & $932 \pm 23$ & $10.22 \pm 0.25$ &
                   $6250 \pm 50$ & $3910 \pm 70$ & $42.88 \pm 0.77$ \\
      & $4\pi$ & $2350 \pm 50$ & $1053 \pm 24$ & $11.55 \pm 0.27$ &
                   $7350 \pm 50$ & $4310 \pm 65$ & $47.30 \pm 0.71$ \\
      & $8\pi$ & $2050 \pm 50$ & $979 \pm 26$ & $10.74 \pm 0.29$ &
                 $7150 \pm 50$ & $3830 \pm 57$ & $42.00 \pm 0.63$ \\
%
%
%
%
\hline
B     & $2\pi$ & $13500 \pm 500$ & $1025 \pm 45$ & $11.23 \pm 0.50$ &
                   $31500 \pm 500$ & $2578 \pm 79$ & $28.28 \pm 0.87$ \\
\cline{6-8}
      & $4\pi$ & $15250 \pm  50$ & $1010 \pm  5$ & $11.08 \pm 0.05$ &
 \multicolumn{3}{c|}{N/A} \\
\hline
C     & $2\pi$ & $99500 \pm 500$ & $1378 \pm  7$ & $15.12 \pm 0.07$ &
 \multicolumn{3}{c|}{N/A} \\
\hline
\end{tabular}
\caption{Time, the number of waves, and nonlinearity parameters at phase transitions.}
\label{tab:transitions}
\end{table*}

Initially, condensate is fed by an inverse cascade carried by an
almost isotropic spectrum of fluctuations.  At this stage, the
nonlinear (interaction) term in the Hamiltonian is less or comparable
to the quadratic term (which describes dispersion for waves or kinetic
energy for particles).  As condensate grows, the first symmetry
breaking appears at $\epsilon\simeq 1$ and the system becomes
anisotropic --- the amplitude and the phase develop patterns on the
scale of the forcing and larger; the spectrum turns into an oval.  The
transformation happens gradually, in the interval $2 \lesssim
\epsilon \lesssim 5$, as the oval slowly becomes thinner at the
waist.

As the condensate amplitude grows further, we observe the sequence of
phase transitions: at first the oval spectrum turns into a dumbbell,
then the symmetry changes from two-fold (two-petal) to three-fold and then to
four-fold, as seen in Figure~\ref{fig:zoo}.  Further transitions are
possible.  While we could not evolve our runs to higher
levels of condensate, we observed other symmetries in the simulations
with different initial conditions (discussed below).
In Figure~\ref{fig:nk_scaled} (left panel) we have shown the regions of
transitions as gray bands in the plot of $\epsilon(t)$.  The
 transitions are accompanied by an increase in the
variance of the over-condensate excitations, $n=N-N_0$, as seen from the
Figure~\ref{fig:nk_scaled}(right panel).

\begin{figure}
\includegraphics[width=0.49\textwidth]{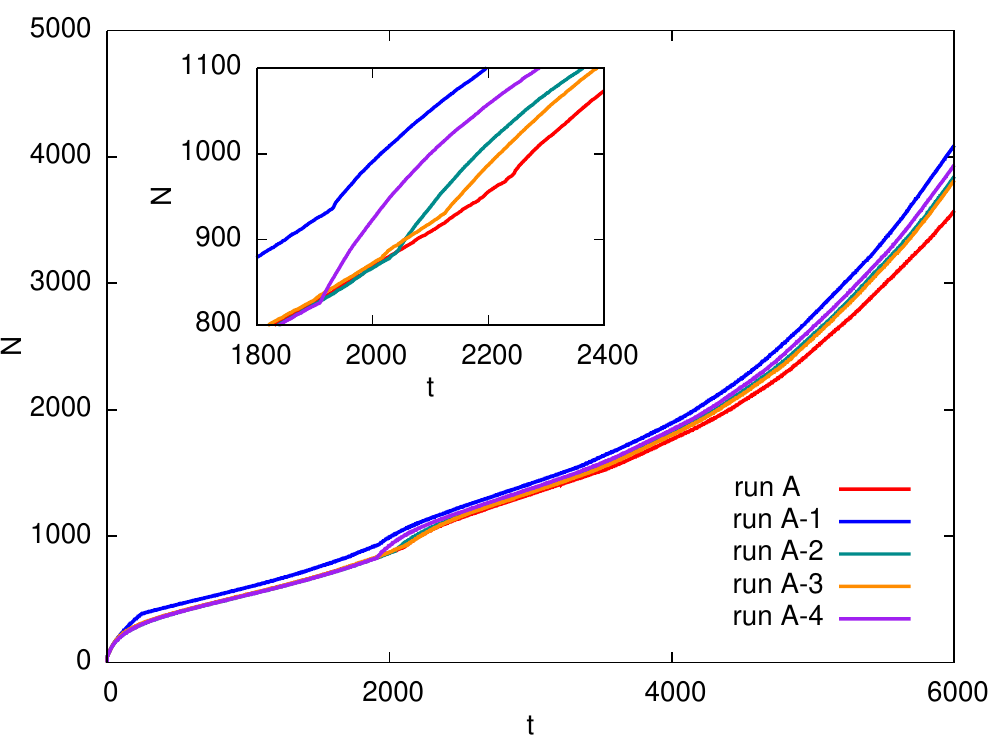}
\caption{
The transition from 2 to 3 petals, seen as the kink in $N(t)$, for
different random phases of the initial condition.  The insert plot is
the zoom to the kink area.  One can see 10\% difference in $N$ at the
moment of transition.  }
\label{fig:random}
\end{figure}

The transition times and condensate amplitudes are listed in
Table~\ref{tab:transitions} for different pumping powers and box
sizes.  The transitions occur sharply yet the threshold values of
$\epsilon$ fluctuate from realization to realization, as we find in the
simulations which only differ in the phases of initial thermal
noise (see Figure~\ref{fig:random}.)  This shows that we are not
dealing with a linear instability with a well-defined threshold.
Indeed, the condensate is stable with respect to the infinitesimal
perturbations. It is likely that our
transitions are of probabilistic nature similar to transitions in a
pipe flow \cite{Hof} or fiber laser \cite{tur} when with the change of a
control parameter (condensate amplitude in our case) one changes the
probability that a finite-amplitude perturbation will lead the system
away to a new state.

\begin{figure*}
\includegraphics[width=0.95\textwidth]{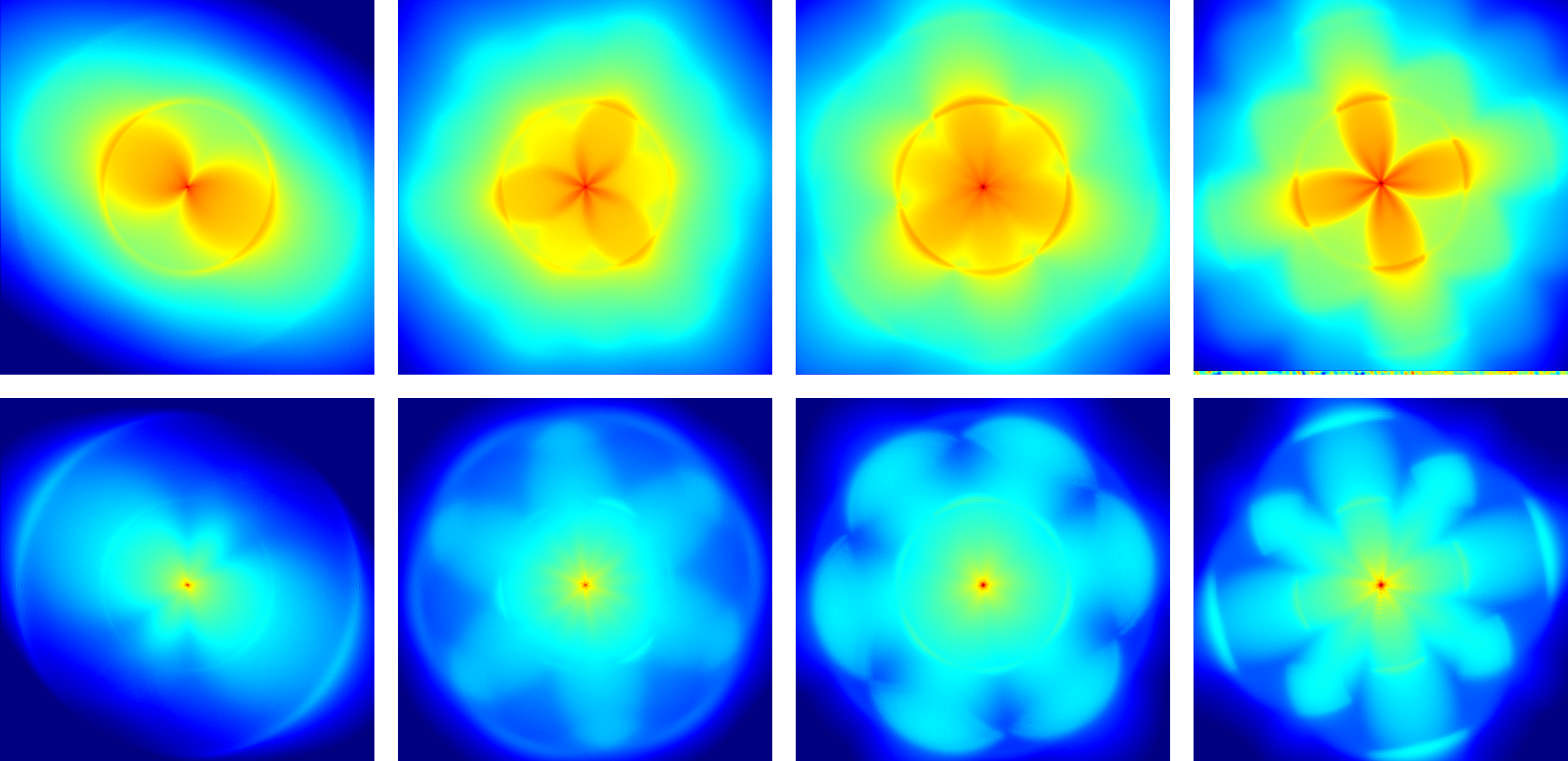}
\caption{Total spectrum (above) and filtered spectrum of the coherent mode
(below) in $L=8\pi$ simulation.  From left to right images correspond
to $t=1500$, $2100$, $6500$ and $7500$ ($N=771$, 1005, 3102, 4202). The averaging was done over 100 time units in stabilized runs.
Colormap is the same as in Figure~\ref{fig:zoo}.
}
\label{img:spectra}
\end{figure*}

\begin{figure*}

\includegraphics[width=0.98\textwidth]{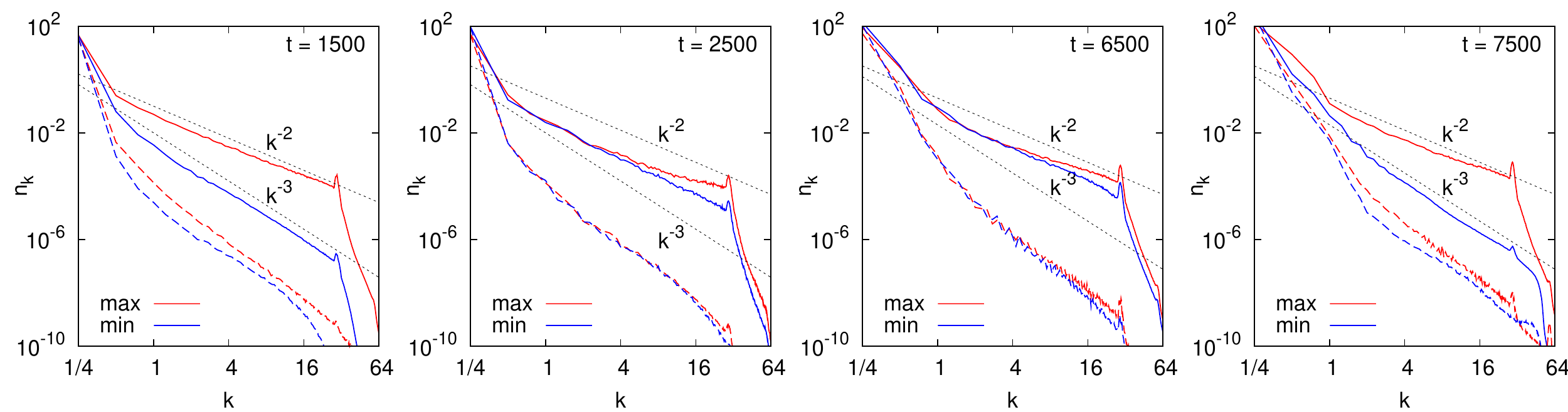}
\includegraphics[width=0.98\textwidth]{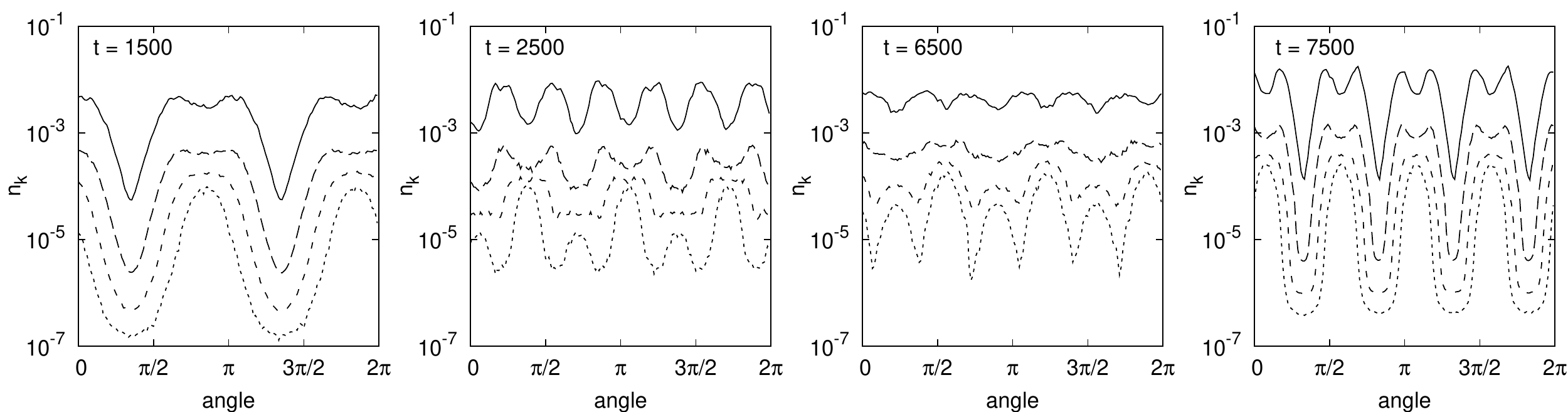}
\caption{
Radial dependency at two directions (first row) and angular dependency at $|k|=8$, 24, 48, 64 (second row, from top to bottom) for the spectra shown in Figure~\ref{img:spectra}. Dashed lines in the radial spectra correspond to the coherent mode.
}
\label{fig:radial}
\label{fig:angular}
\end{figure*}

It can be
recognized in the second, third, and fourth panels of the second row in
Figure~\ref{fig:angular} that the maxima at high $k$ correspond to
minima at low $k$, in agreement with the waveguide argument presented in Section~\ref{sec:aniz}.
However, the symmetry of the spectra both at low and high
$k$ is definitely not that of a box.
The minima and maxima are directed at different angles and the
spectrum of turbulence can have three-fold symmetry that is not that
of a square box.
Bottom row of Figure \ref{img:spectra} shows that the mode coherently oscillating
with the frequency $N$ has an intricate spatial structure and
multiscale correlations the anisotropy of which is directly related to the
anisotropy of over-condensate fluctuations.
The spectrum of the coherent mode has the doubled number of maxima and is generally more symmetric than the whole
spectrum, as seen from Figure~\ref{img:spectra}.
For nonzero momenta, over-condensate fluctuations
dominate over the coherent part.

\begin{figure}
\includegraphics[width=0.49\textwidth]{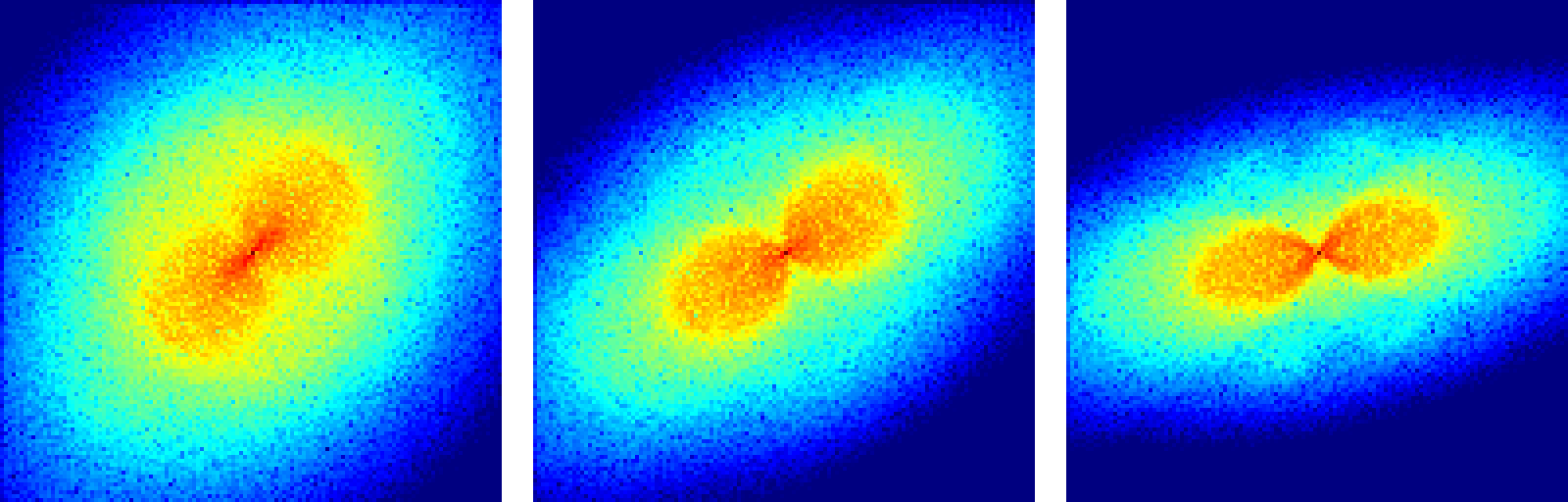}
\caption{
Form of the spectrum at $t=10000, 40000, 80000$, from left to right,  in Run C at $L=2\pi$.
The spectrum angular width decreases as $N$ grows.}
\label{fig:oval}
\end{figure}

\begin{figure}
\includegraphics[width=0.40\textwidth]{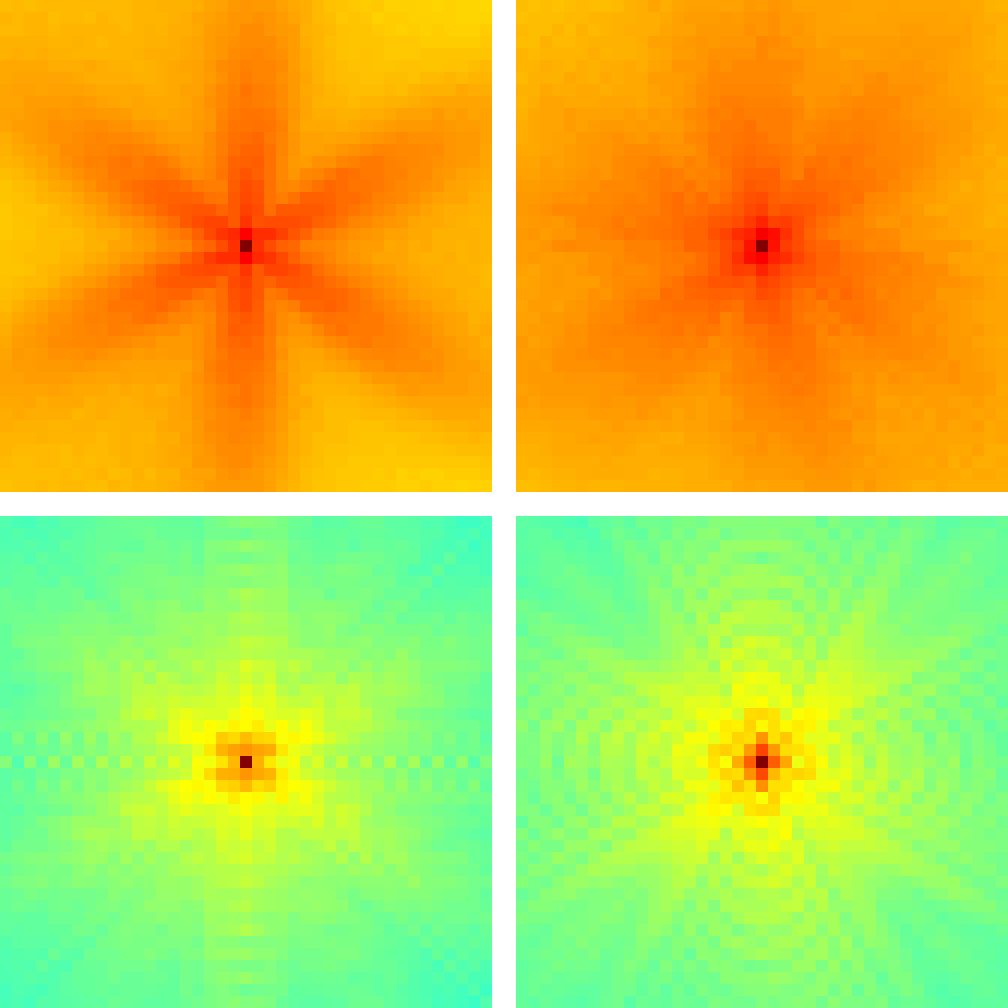}
\caption{Zoom into the central part of the total spectra (above)
and the spectra for the coherent mode (below) for $L=4\pi$
and $t=3000,6000$; both correspond to three-fold symmetry.
One can see the four-fold symmetry appearing at low $k$
in the lower right panel.}
\label{fig:zoom}
\end{figure}

In addition to phase transitions between different symmetry states,
the spectra undergo slow evolution between transitions.  The
intensity fluctuates, the whole structure may slowly rotate or
swing.  With time, two-petal spectra concentrate in a more narrow
angle, as seen in Figure~\ref{fig:oval}, and broaden again just before
the transition.  The three-petal spectra broaden with time, especially
at high $k$. The third columns in  Figures~\ref{img:spectra},\ref{fig:angular} show spectra just before the transition, when they become close to six-petal; the seeds of the new (four-fold) symmetry visible at low $k$ in the coherent part, as seen in Figure~\ref{fig:zoom}.  The population of
the pumping shell increases between the transitions and falls during the transitions, which explains a complicated character of the evolution of the total number of waves.

At any moment, except possibly very close to the transition events, and for
all symmetries, the angular width of the spectra decreases towards
larger wavenumbers, as seen in
Figures~\ref{img:spectra},~\ref{fig:angular}.  This may be analogous to
spectrum anisotropization observed for acoustic turbulence
\cite{ZLF,FM}. It is worth stressing the difference in two cases: acoustic
turbulence studied there corresponds to a direct energy cascade
(realized by pumping-generated long waves turning into shocks),
while here we have an inverse cascade. One can therefore say that as
an inverse cascade proceeds towards larger wavelength, the spectra are
getting wider as weak turbulence theory would predict \cite{ZLF}; yet
presently we see no way it can predict the spontaneous appearance of anisotropy
and changes in symmetry, particularly, significant anisotropy at the
pumping scale.

One may suggest a possible physical mechanism of phase transitions as
follows: phonons effectively interact only within the angle
$k/\sqrt{N_0}$, which decreases as $N_0$ grows, so the turbulence
tends to be broken into jets.  The number of jets $j$ increases
proportionally to the inverse of the interaction angle i.e. as
$N_0^{1/2}\propto\epsilon^{1/2}$.  Indeed, we observe that the
transitions happen approximately at $\epsilon_j \sim
j^2$.  Even if this is indeed the basic mechanism of
the transitions, we still lack any understanding of whether there is a
unifying principle that can predict which turbulent state is realized
the way variational principle does it for thermal equilibrium. One direction worth exploring is whether one can develop
an approach similar to the weak crystallization theory \cite{KLM}
despite the fact the we are dealing with developed turbulence and power-law
spectra carrying flux in ${\bf k}$-space, as seen in Figs~\ref{fig:radial},
\ref{fig:nkflow} below.

Finally, we made separate runs taking initial conditions for $\psi$ as
a thermal noise plus large constant (pre-existing condensate). In
these cases, the same states (with 2,3,4 petals) appear i.e. they are
true attractors, independent of the history.  We also observed 6-petal
spectra with triangular spatial pattern in the amplitude, and 8-petal
spectra with the spatial
distribution comprised of patches of square patterns oriented at
45 degree angle. The 6-petal spectra was observed at higher
$\epsilon$ than 4-petal spectra, although pre-existing condensate
makes all thresholds lower and possibly less sensitive to the phases of the
initial thermal noise.  We speculate that this is because in the runs
with a pre-existing condensate, the spectrum first develops a uniformly
populated pumping shell which is then free to break apart into the appropriate
number of jets, contrary to evolving systems where the previous
configuration has to re-populate the pumping shell.  The ability of
the system to quickly find its equilibrium state opens horizons for a
much more extensive exploration of parameter space, which hopefully
will be a subject of future studies.

\subsection{Orientational order}

\begin{figure}
\includegraphics[height=56mm]{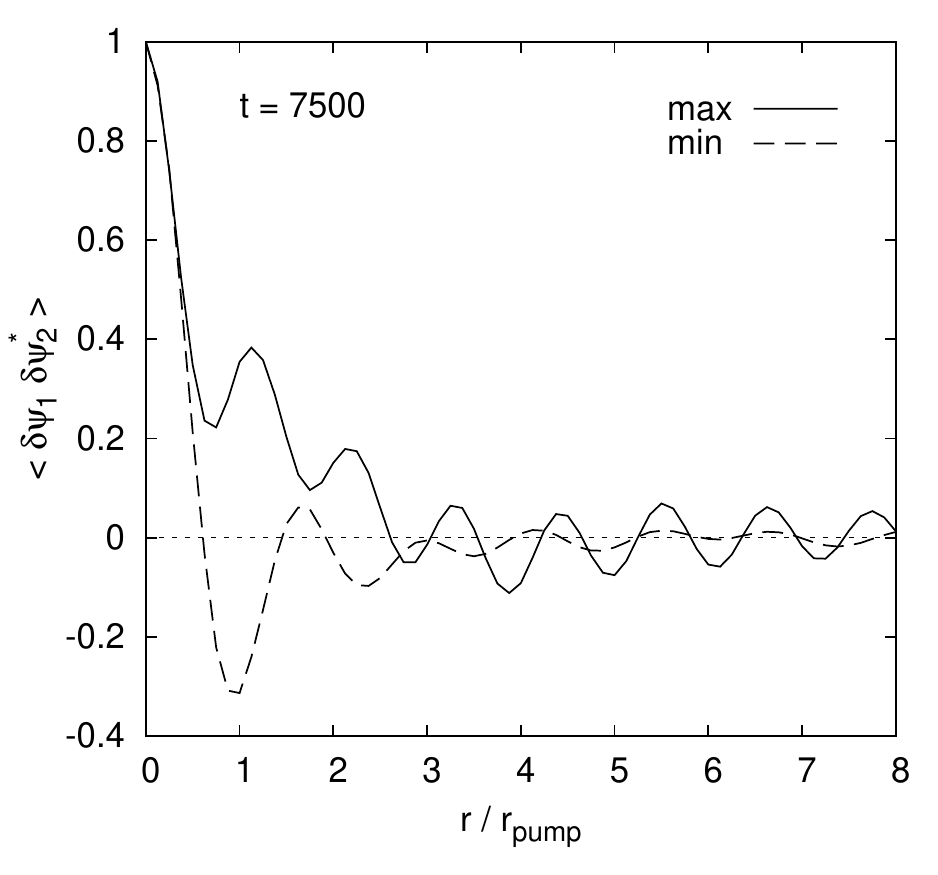}
\caption{Correlation function of the amplitude of the
over-condensate fluctuations at $t=7500$ for $L=2\pi$.
}
\label{fig:Corfun}
\end{figure}

For two-fold and three-fold spectra, turbulence statistics is
translational invariant. The most remarkable finding is seen at the
last  image  of the top row in Figure~\ref{fig:zoo}: on top of a
long-range  orientational order, a short-range
positional order appears. This state is between the solid and the
isotropic liquid very much like hexatic phase in 2d melting
\cite{Hex}. It requires future detailed studies to establish whether
this is indeed a turbulent analog of the
Berezinski-Kosterlitz-Thouless transition; to avoid misunderstanding,
let us stress that there are no vortices (holes) in the
condensate at this stage, the phase fluctuates weakly; it is the
small condensate perturbations which are getting ordered. These
spatial correlations are also seen in Figure~\ref{fig:Corfun} which
shows the correlation function of the amplitude, $\overline{\psi({\bm
x})\psi({\bm x}+{\bm r})}-\overline{\psi}^2$, where overline means
averaging over the position ${\bm x}$.

The type of crystallization that we observe
is very much different from the vortex crystallization with a
wavelength externally imposed by a cut-off in 2d incompressible
turbulence \cite{SY}.

\subsection{Anomalous correlations and collective oscillations}

As we have predicted above (Section \ref{sec:anom-cor}) anomalous correlations must exist between any two contra-propagating waves. In our direct numerical simulations we have confirmed that this is true at least for the two lowest modes of
the box, their total phase is indeed shifted by $\pi$ relative to the
condensate. These modes give the main contribution of the
collective oscillations of the system which are periodic conversions
of waves from condensate to over-condensate with the total number of
waves unchanged.

We checked that in the numerical data the steady-state anomalous
correlation function indeed has the phase shifted by $\pi$ relative to
the condensate value squared. For example, for $L=4\pi$ and $N=570$,
the modes that have the largest amplitudes, with $(k_x,k_y)=(1,0),
(1,1), (1,-1)$, have their phase differences
$(\phi_k+\phi_{-k}-2\phi_0)/\pi=1.01,1.02,0.99$ respectively.

It was noticed in~\cite{Fal84} that when broad turbulent spectra
coexist with a sharp spectral peak these two groups of waves generally
have different relaxation times which makes possible oscillations
about a steady state.  For three-wave interaction, an integral model
(of a predator-prey type) that describes the evolution of the total
numbers of waves in the two groups has the form: $dN_0/dt=-b N_0+N_0n$,
$dn/dt=\gamma n - N_0n$, which gives the oscillations with the frequency
$\sqrt{\gamma b}=\sqrt{\bar n\bar{N_0}}$~\cite{Fal84}.

The oscillations for NSE with a condensate were reported
in~\cite{Opt} where a similar model was suggested: $dN_0/dt=-b
N_0 + N_0 n^2$, $dn/dt=\gamma n - N_0^2 n$. This model gives the
steady-state values $\bar{N_0} = \sqrt{\gamma}$, $\bar n=\sqrt{b}$ and
the same frequency of oscillations $\sqrt{\gamma b}=\sqrt{\bar n \bar
{N_0}}$.  The most evident defect of the latter model is that it does not
conserve the total number of waves, moreover, the steady-state values
are wrong too (in reality, $\bar N_0\propto 1/b$ while $\bar n$ is
practically independent of $b$ in wide intervals).
We notice here that to describe the collective oscillations of the turbulence-condensate system, one needs to account properly for the anomalous correlation function, as was done while deriving  system (\ref{phase2}). In particular, our suggested system does describe the oscillations around a steady state. Consider, $N_0=\bar N_0+M$, $n_k=\bar n_k+m$,   assuming that the steady-state values satisfy $\bar N_0 \gg \bar n_k$. We then get
\begin{figure*}
\includegraphics[width=0.9\textwidth]{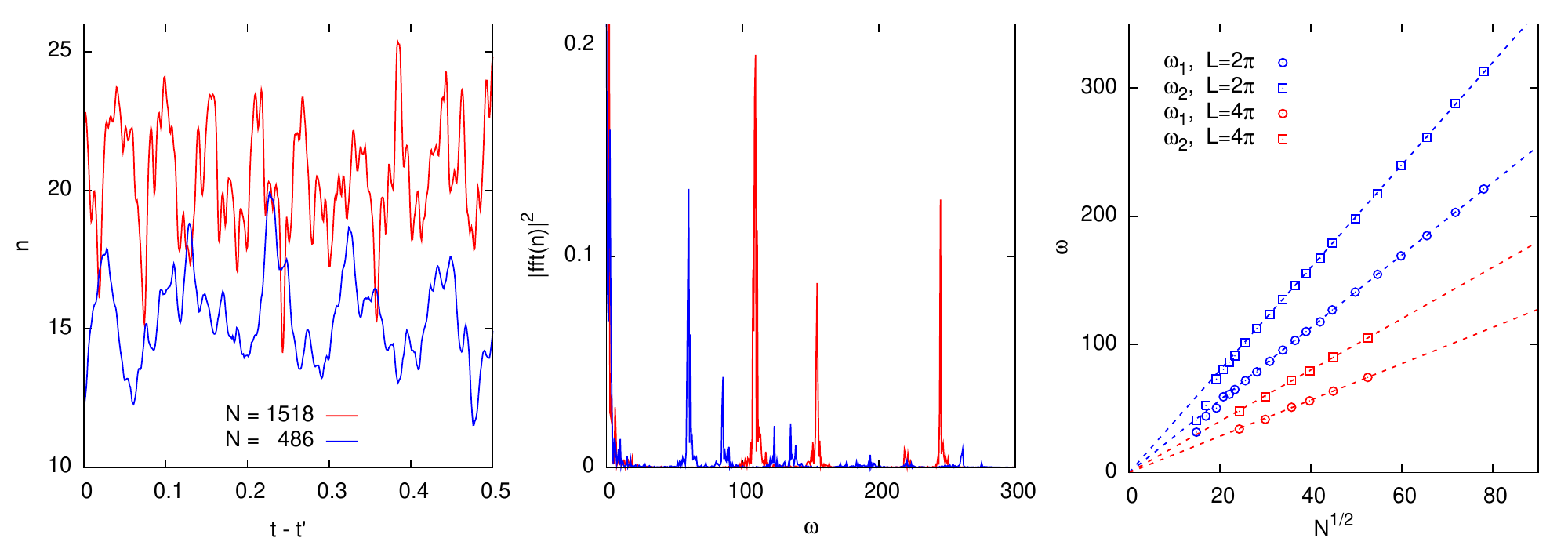}
\caption{
Left: oscillations in over-condensate in Run A, $L=2\pi$ box, zoomed
around times $t'=800$ ($N=486$) and $t'= 3500$ ($N=1518$).  Middle:
frequency spectrum of the time-dependence of the number of
over-condensate waves for the same run in time intervals $[800,810]$
and $[3500,3510]$.  Right: first two frequencies as function of the
number of waves in $L=2\pi$ and $L=4\pi$ boxes.  Dashed lines
have the slopes 4, $2\sqrt{2}$, 2, $\sqrt{2}$.  }
\label{fig:oscil}
\end{figure*}

\begin{eqnarray}
\begin{split}
{dm\over dt} &=-b=-{dM\over 2dt}\,, \nonumber\\
{dB\over dt}  &=-4(\bar N_0+k^2)C-4\bar N_0^2m \,, \nonumber\\
{dC\over dt}  &= (\bar N_0+k^2)B
\end{split}
\label{osc1}
\end{eqnarray}
\begin{figure}
\includegraphics[width=0.48\textwidth]{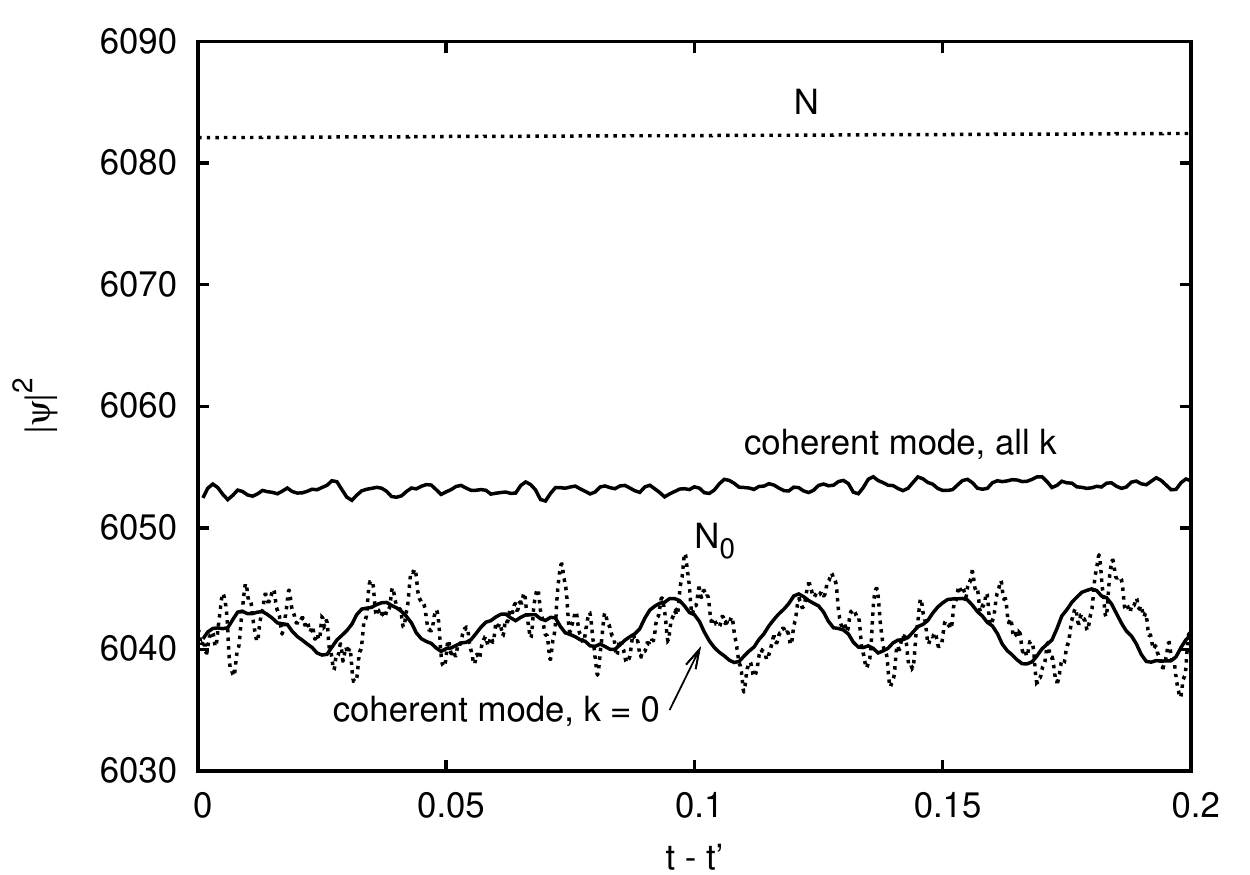}
\caption{Oscillations of number of waves
in the condensate, $N_0$,  and in the coherent mode for $L=2\pi$.}
\label{fig:oscil1}
\end{figure}

The system above describes collective oscillations with twice the Bogolyubov frequency: $M,m,B,C\propto \exp(-i\,\Omega t)$, $\Omega=2\Omega_k=2\sqrt{2\bar N_0k^2+k^4}$.
Indeed, numerical simulations show
that the oscillations have frequencies that are equal to twice that of the lowest modes of the
Bogolyubov spectrum. For the box with $L=2\pi$ we find that the
modulus of the anomalous correlation functions
$\sigma(k_x,k_y)=\sigma(1,0)$ and $\sigma(1,1)$ oscillate respectively
with the frequencies $2\sqrt{2 N_0}$ and $4\sqrt{N_0}$.  These
frequencies (which are much smaller than the frequency of the
condensate $N$) essentially do not depend on the level of
over-condensate fluctuations, they are clearly seen in the oscillations of
the condensate amplitude and of the normal correlation functions shown in
Figure~\ref{fig:oscil}.  As seen from Figure~\ref{fig:oscil1}, the
total number of waves is unchanged. The number of over condensate
waves (about $40=6080-6040$) consists of a coherent part (about
$10=6050-6040$), the rest (about 30) are waves with other frequencies. Note
that the number of over-condensate waves exceeds substantially the
amplitude of the oscillations (about 5). The oscillations can be thus
treated in a linearized approximation.

\subsection{Moments, spectra and flux lines}
Table~\ref{tab:stats} shows the integral statistical characteristics of over-condensate fluctuations for different times and levels of the condensate. We define $\delta\psi=\psi-\bar\psi$ and compute the spatial averages of its moments.
The fourth column gives the second moment, which is the fraction of waves with nonzero wavenumbers: $ \overline{|\delta\psi|^2}/N_0=(N-N_0)/N_0$. The last column shows the
flatness, which is close to the Gaussian value 3.
Far from the transitions, the flatness is actually somewhat lower
than  3 and decreases as the condensate grows, i.e. the condensate effectively suppresses
strong over-condensate fluctuations contrary to what condensate vortices do to 2D
incompressible turbulence \cite{Shats}. Only for $t=6500$, which is close to the three-to-four transition, we find flatness slightly exceeding 3, i.e. somewhat stronger fluctuations. Note in passing that the coherent mode at all scales has substantially smaller
flatness, which means that  strong fluctuations of the mode are
even more rare than for a Gaussian random quantity.

\begin{table}
\begin{tabular}{| c | r | r | c | c |}
\hline
 \quad $t\quad$ &\quad  $N\quad$ &\quad $N_0\quad$ & \  $\;\overline{|\delta\psi|^2}/N_0\;$\  &
$\;\overline{\,|\delta\psi|^4} / \left( \overline{\,|\delta\psi|^2 }\right)^2$ \\
\hline
1500 &  771 &  751 & 0.027 & 2.95 \\
2500 & 1166 & 1133 & 0.029 & 2.67 \\
6500 & 3102 & 2871 & 0.080 & 3.17 \\
7500 & 4200 & 3986 & 0.054 & 2.40 \\
\hline
\end{tabular}
\caption{Integral statistics of fluctuations corresponding to spectra in Figures~\ref{img:spectra},\ref{fig:angular}.}
\label{tab:stats}
\end{table}

Let us stress that despite anomalous correlations, symmetry-breaking
phase transitions and collective oscillations (all properties normally
associated with narrow spectra in ${\bf k}$-space), our spectra are
wide along and across the jets, decaying approximately by power laws, as seen
in Figure~\ref{fig:radial}. Note that $1/k^2$ is the spectrum of weak
turbulence inverse cascade which coincides in this case (up to a
logarithmic factor) with the thermal equilibrium, while $1/k^3$ is the
spectrum due to the shock waves.

\begin{figure}
\includegraphics[width=0.49\textwidth]{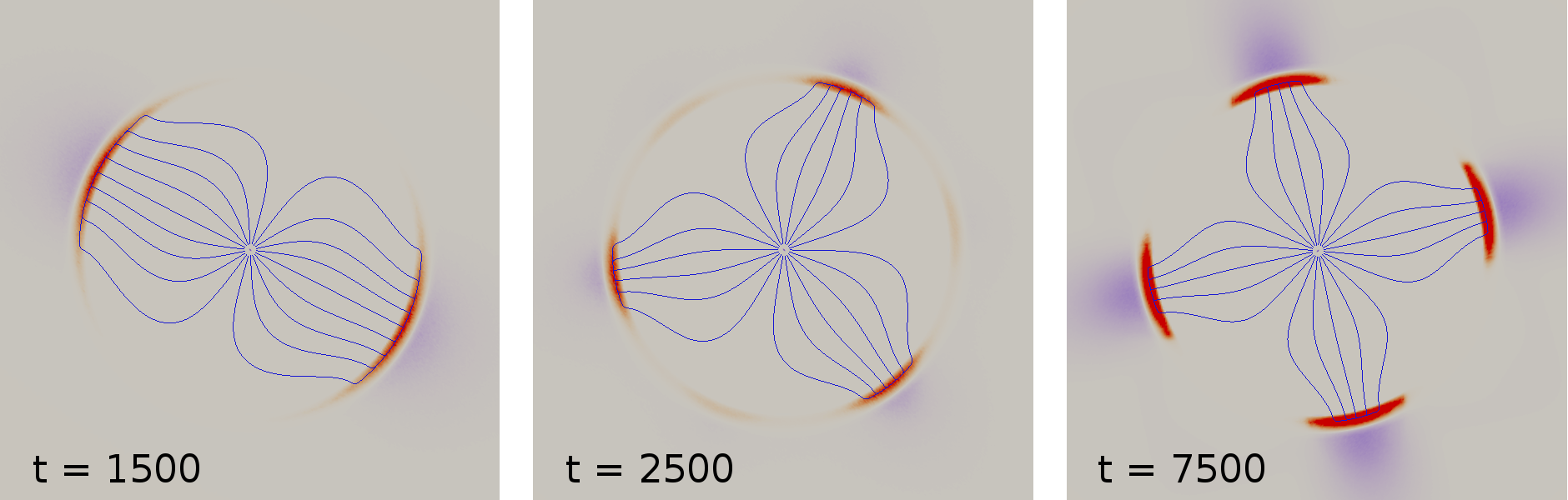}
\caption{
Source distribution and flux flow lines at $t=1500$, $t=2500$, and $t=7500$ for $L=8\pi$.}
\label{fig:nkflow}
\end{figure}

Let us now describe the lines of the
flux that these spectra carry towards the condensate.  Conservation of
waves allows one to write NSE as a continuity equation and define the
flux, ${\bf J}_k$, via the relation   ${\rm div}{\bf J}_k=\langle |\psi_k|^2\rangle \, \gamma_k
\equiv Q_k$.
In this Poisson equation, $Q_k$ plays the role of a charge
density and the flux plays the role of potential.
We compute $Q_k$
and solve the above equation.  Figure~\ref{fig:nkflow} shows the
flux lines while positive/negative source $Q_k$ is shown by red/blue
regions.

\subsection{Non-universality of the inverse cascade turbulence}

\begin{figure}
\includegraphics[width=0.4\textwidth]{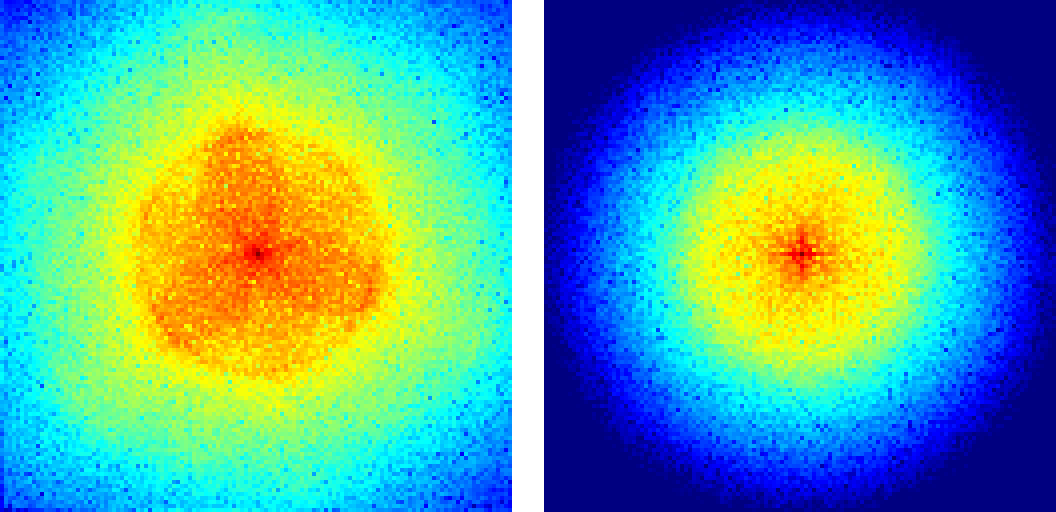}
\caption{
Spectra for an instability (left) and the random force (right)   at
about the same levels of condensate for $L=2\pi$. One sees that force-driven
turbulence remains isotropic except for a few box-size modes, while the
instability-driven turbulence develops a three-petal spectrum in the whole of
${\bf k}$-space.}
\label{fig:compar}
\end{figure}

It is important that the phase transitions happen only when the waves
are excited by an instability. If we excite waves by an additive
random force concentrated in the same ring in ${\bf k}$-space with
about the same amplitude, turbulence remains close to isotropic at all
wavenumbers exceeding unity and at all condensate levels, as seen in
Fig~\ref{fig:compar}. To the best of our knowledge, this is the first demonstration
of significant non-universality of turbulence with respect to excitation mechanism. It is particularly striking that we find it for an inverse cascade which is generally particularly robust \cite{sym}. Apparently, the condensate-turbulence interaction can make turbulence non-universal.

 This set of simulations was performed for $L=2\pi$ and
the following procedure was used. First a run with an
instability excitation was done  with the initial condition  as a large constant
(corresponding to $N_0=1000$) plus small thermal noise. The run was stopped
after it produced a 3-petal spectrum   shown in the left panel of
Fig.~\ref{fig:compar}; during that time the total number of waves grew
only by 3.6\% from its initial value.  Second, a run with an additive
forcing was performed. In this run  the
multiplicative term $\gamma_p(k) \psi_k$ was replaced by an
additive forcing $f({\bf k},t)$ (the multiplicative damping
term, $\gamma_d(k)$, was preserved). This forcing term was non-zero only inside the
pumping shell $k_l \leq k \leq k_r$, it had random phases uniformly
distributed in the interval $[0,2\pi]$ and uncorrelated both in ${\bf
k}$-space and time. Its amplitude was chosen to be constant $|f({\bf
k},t))|=f$ the value of which was determined by a condition that both
additive and multiplicative forcing have the same average amplitude in
the $\bf{k}$-space: $$ f=\frac{1}{\pi(k_r^2-k_l^2)} \,\int_{k_l \leq k
\leq k_r} \, \gamma_p(k) \, |\psi_k| \, d^2 k\ .$$

\smallskip

To conclude, discerning guiding principles that govern the physics far
from equilibrium remain elusive. Is there any quantity (similar to entropy) that the
system tries to optimize by undergoing the series of phase transitions
described here?

We thank A. Zamolodchikov, V. Lebedev, E. Kats, A. Finkelstein and
B. Shraiman for helpful discussions. This research was supported by
the grants of the BSF, ISF and by the Minerva Foundation funded by the
German Ministry for education and research. Part of the work was done
at KITP where it was supported by the National Science Foundation
under Grant NSF PHYS-51164. Computations were done at the New Mexico
Computing Application Center and at UNM Center for Advanced Research Computing.


\end{document}